\documentstyle[seceq,twocolumn]{jpsj}
\def\sbox#1{\mbox{\scriptsize #1}}
\def\H{{\cal H}}
\def\img{{\mbox{i}}}
\def\mub{{\mu_{\sbox{B}}}}
\def\H{{\cal H}}

\def\v#1{\mib #1}
\def\simleq{\mbox{\raisebox{-1.0ex}{$\stackrel{<}{\sim}$}}}

\def\ln{\mbox{ln}}

\def\runtitle{Interacting Boson Theory of the F-AF Alternating Heisenberg Chain
}

\def\runauthor{Kazuo {\sc Hida}}

\title
{
Interacting Boson Theory of the Magnetization Process of the Spin-1/2 Ferromagnetic-Antiferromagnetic Alternating Heisenberg Chain
}

\author
{Kazuo {\sc Hida}
\footnote{e-mail: hida@riron.ged.saitama-u.ac.jp}}

\inst
{
Department of Physics, Faculty of Science,\\ Saitama University, Urawa, Saitama 338
}

\recdate
{
October 31, 1997
}

\abst
{
The low temperature magnetization process of the ferromagnetic-antiferromagnetic Heisenberg chain is studied using the interacting boson approximation. In the low field regime and near the saturation field, the spin wave excitations are approximated by the $\delta$ function boson gas for which the Bethe ansatz solution is available. The finite temperature properties are calculated by solving the integral equation numerically. The comparison is made with Monte Carlo calculation and the limit of the applicability of the present approximation is discussed.}

\kword
{
ferromagnetic-antiferromagnetic alternating Heisenberg antiferromagnet, spin wave, Bethe ansatz, magnetization
}

\begin{document}
\sloppy
\maketitle

%\draft

\section{Introduction}
The spin-1/2 ferromagnetic-antiferromagnetic alternating Heisenberg chain has been intensively studied related with the Haldane gap problem\cite{kh1,khd,khr,sn1,hag1,yam1,tak1,sak1}. In the limit of strong ferromagnetic bond, this model tends to the spin-1 antiferromagnetic Heisenberg chain(AFHC). On the other hand, it reduces to an assembly of trivial isolated dimers if the ferromagnetic interaction vanishes. It also tends to the dimerized spin-1/2 AFHC which describes the spin-Peierls state if the ferromagnetic bonds are replaced by the antiferromagnetic bonds. Thus the ground state of this model interpolates the Haldane gap state and the spin-Peierls state continuously.  

The magnetization process of this model has been also studied theoretically and experimentally\cite{sn1,hag1,yam1,tak1,sak1}. It is pointed out that the critical behaviors at the lower critical field and the saturation field are described by the free fermion theory, which is equivalent to the boson with hard core interaction\cite{sak1}. It is the purpose of the present work to investigate the thermal effect on the magnetization process of this model. 

The Hamiltonian of the present model is given by
\begin{equation}
\label{eq:ham1}
\H =J\sum_{l=1}^{N}\v{S}_{2l} \v{S}_{2l+1} + J'\sum_{l=1}^{N}\v{S}_{2l-1} \v{S}_{2l} -g\mub H\sum_{l=1}^{2N}S_{l}^{z},
\end{equation} 
where $\v{S_l}$ is the spin operator with magnitude $1/2$ on the $l$-th site. It is assumed that $J$ is positive (antiferromagnetic) and $J'$ is negative (ferromagnetic).  The external magnetic field, the electronic $g$ factor and the Bohr magneton are denoted by $H, g$ and $\mub$ , respectively. In the following, we take the unit $J=1$ and $g\mub=1$. The ratio $J'/J$ is denoted by $\beta$. In the next section, the magnetization process of this model at low temperature $T << 1$ is studied by means of the mapping onto the interacting boson system which is solvable by the Bethe Ansatz method. The quantum Monte Carlo results are presented in \S 3. In the last section, the both results are compared and the limit of applicability of the interacting boson approximation is discussed.

\section{Interacting Boson Approximation}
\subsection{Near the saturation field}

In the vicinity of the saturation field, we may apply the spin wave method starting from the fully polarized state. In order to incorporate the nonlinear terms in a convenient way, we apply the Dyson-Maleev transformation\cite{ts1,fjd1,svm1} as follows, 

\begin{equation}
S_l^{z}=(\frac{1}{2}-a_l^{\dagger}a_l),\ S_l^{+}=(1-a_l^{\dagger}a_l)a_l,\ S_l^{-}=a_l^{\dagger}, 
\end{equation}
to obtain the Boson representation of the Hamiltonian as
\begin{eqnarray}
\label{eq:hamb1}
\H &=& \H_0 + \H_{\mbox{int}},\\
\H_0 &=&\sum_{l=1}^{N}\{\frac{1}{4} - \frac{1}{2}(a^{\dagger}_{2l} -a^{\dagger}_{2l+1})(a_{2l} - a_{2l+1})\} \nonumber \\
&+& \beta \sum_{l=1}^{N}\{\frac{1}{4} - \frac{1}{2}(a^{\dagger}_{2l-1} -a^{\dagger}_{2l})(a_{2l-1} - a_{2l})\}, \\
\H_{\mbox{int}}&=&
- \frac{1}{2} \sum_{l=1}^{N} a^{\dagger}_{2l}a^{\dagger}_{2l+1}(a_{2l}-a_{2l+1})^2 \nonumber \\
  &-& \frac{\beta}{2} \sum_{l=1}^{N} a^{\dagger}_{2l-1}a^{\dagger}_{2l}(a_{2l-1}-a_{2l})^2. 
\end{eqnarray} 

By the Fourier transformation 

\begin{eqnarray}
a_{2l} &=& \frac{1}{\sqrt{N}}\sum_{k}c_k \exp (\img k l),
\nonumber \\
a_{2l+1} &=& \frac{1}{\sqrt{N}}\sum_{k}d_k \exp (\img k (l+\frac{1}{2})),
\end{eqnarray}
where the length of the unit cell (=twice lattice constant) is set equal to unity, $\H_0$ is transformed as,

\begin{eqnarray}
\H_0 &=& E_0 + \sum_{k}\left\{\left(H-\frac{1+\beta}{2}\right)(c^{\dagger}_k c_k + d^{\dagger}_k d_k) 
\right.  \nonumber \\ &+& \left.\frac{1}{2}(J(k)d^{\dagger}_k c_k + J^*(k) c^{\dagger}_k d_k ) \right\},
\end{eqnarray}
where 
\begin{equation}
E_0  = \frac{(1+\beta)N}{4}-NH,\ \ J(k) = \mbox{e}^{\img k/2}+\beta \mbox{e}^{-\img k/2}. 
\end{equation}
After the Bogoliubov transformation which diagonalizes $\H_0$, we have

\begin{equation}
\H_0 = E_0 + \sum_{k}\{ E_{+}(k)\alpha^{\dagger}_k \alpha_k + E_{-}(k)\beta^{\dagger}_k \beta_k \},
\end{equation}
where
\begin{equation}
\label{eq:bosh}
E_{\pm}(k)=-\frac{1}{2}\left(1+\beta\pm \sqrt{1+\beta^2 + 2\beta \cos k} \right)+ H,
\end{equation}
and $ \alpha_k$ and $\beta_k$ are boson operators given by the linear combinations of $c_k$ and $d_k$.
The magnetization per unit cell $M$ is expressed as 
\begin{equation}
M = 1 - \frac{1}{N}\sum_{k} \{<\alpha^{\dagger}_k \alpha_k> +<\beta^{\dagger}_k \beta_k> \}.
\end{equation}
where $< >$ represents the thermal average. The number of the quasiparticles corresponds to the number of inverted spins. For $H \geq 1$, the excitation energy of the quasiparticles are positive corresponding to the fully magnetized ground state, while for  $H < 1$, the quasiparticles with momenta $k \simeq \pi$ have negative energy and these quasiparticles are present even at zero temperature leading to the deviation from the saturated magnetization. Therefore the saturation field $H_{\mbox{s}}$ is given by $H_{\mbox{s}}=1$. Near the saturation field at low temperatures, the density of the quasiparticles is low  and  they are distributed around the minimum of the excitation spectrum at $k \sim \pi$ of $E_{+}$. Therefore we only take into account the $+$ branch and shift the origin of the momentum by $\pi$ as $k=\pi+q$. In the following, we use $q$'s instead of $k$'s as momentum indices. Expanding around $q=0$ and keeping the important terms at $k \sim \pi$, we find
\begin{eqnarray}
\label{eq:mfham1}
\H &=& E_0 + \sum_{q}\{-1 + H +\frac{\xi^2q^2}{2}\}\alpha^{\dagger}_{q} \alpha_{q}\nonumber \\
  &+& \frac{1}{2N} \sum_{q_1,q_2,q_3,q_4} \alpha^{\dagger}_{q_1}  \alpha^{\dagger}_{q_2}  \alpha_{q_3}  \alpha_{q_4} \delta_{q_1+q_2, q_3+q_4}. 
\end{eqnarray}
Here the same Fourier and Bogoliubov transformations are applied to the interaction Hamiltonian $\H_{\mbox{int}}$ and the terms containing the $-$ branch are neglected. The coefficient of $\H_{\mbox{int}}$ is independent of $\beta$ at $k \sim \pi$  as far as $\beta < 0$. This is due to the fact that the $+$ branch represents the in-phase motion of the spins connected by the ferromagnetic bonds. The correlation length $\xi$ is given by $\xi = \sqrt{\mid  \beta\mid /2(1 + \mid \beta \mid)}$. The Hamiltonian (\ref{eq:mfham1}) yields the one for the $S=1$ AFHC \cite{ts1} in the limit $\beta \rightarrow -\infty$. This is the Hamiltonian of the $\delta$-function boson which is solved by the Bethe Anzatz method by Lieb and Liniger\cite{ll1}. According to  Yang and Yang\cite{yy1}, the finite temperature properties of this model is described in terms of the solution of the integral equations,
\begin{equation}
\label{eq:eps}
\epsilon(q)=-A+q^2-\frac{tc}{\pi}\int_{-\infty}^{\infty}\frac{dq'}{c^2+(q-q')^2}\ln(1+\mbox{e}^{-\epsilon(q')/t}),
\end{equation}
\begin{equation}
\label{eq:rho}
2\pi\rho(q)(1+\exp (\epsilon(q)/t))=1+2c \int_{-\infty}^{\infty}\frac{\rho(q')dq'}{c^2+(q-q')^2},
\end{equation}
where
\begin{equation}
A=\frac{2(1-H)}{\xi^2},\ \ \ c=\frac{1}{\xi^2},\ \ \ t=\frac{2T}{\xi^2}.
\end{equation}

The magnetization per unit cell is given by
\begin{equation}
M = 1 - \int_{-\infty}^{\infty}\rho(q)dq.
\end{equation}
This set of integral equations are solved numerically to obtain the magnetization curves.

\subsection{The low field regime} 

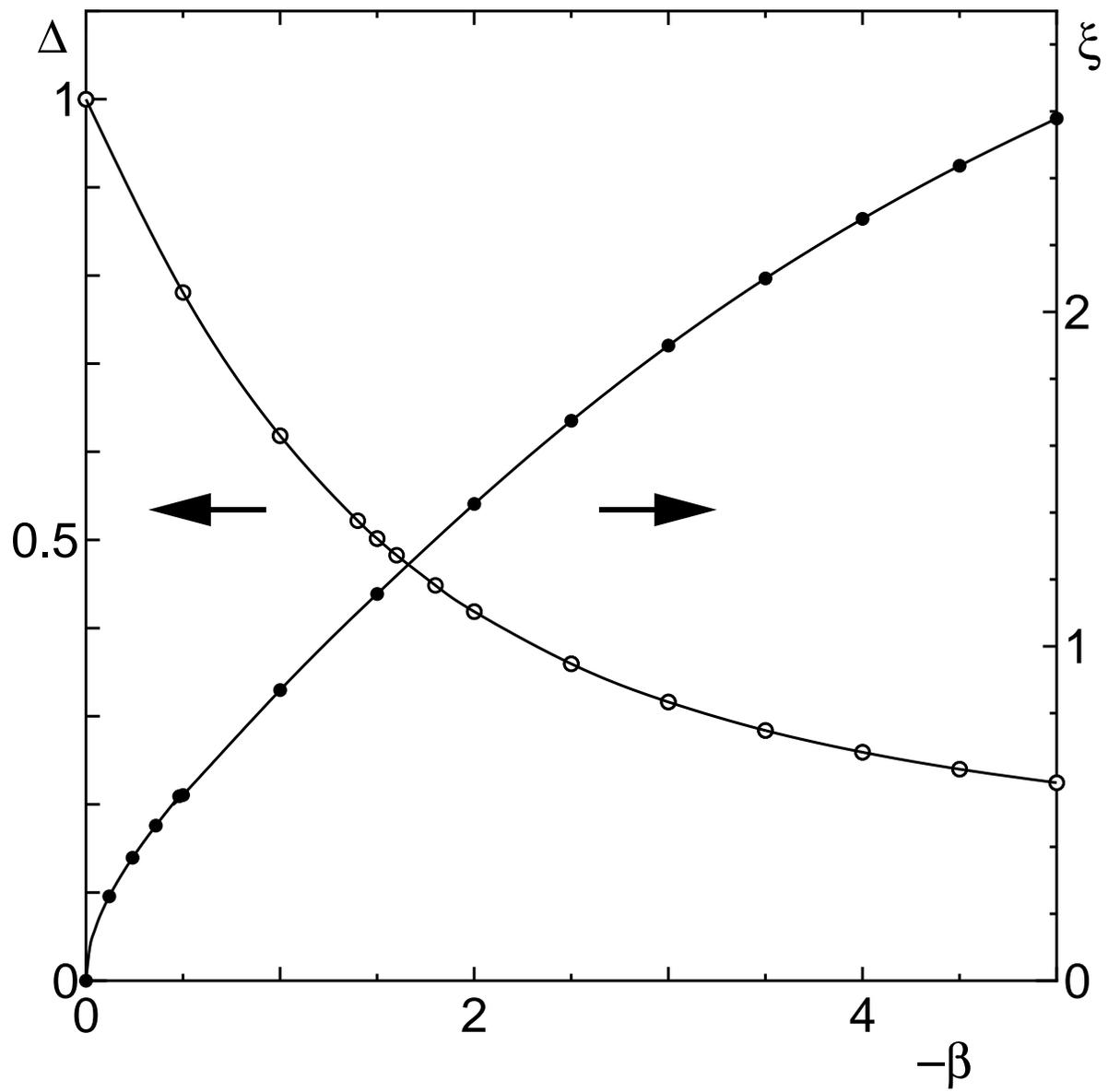
\begin{figure}
\caption{The $\beta$-dependence of $\Delta$ (open circles) and $\xi$ (filled circles) calculated from the exact diagonalization data in ref \citen{khd}. The solid lines are guides for the eye.}
\label{fig1}
\end{figure}

In the low field regime $0 < H \sim \Delta \equiv H_{\mbox{c}}$, where $\Delta$ is the energy gap, the excitations from the ground state are assumed to be the spin-1 bosons with parabolic dispersion above the gap $\Delta$ at $k=\pi$ and short range repulsion similarly to the spin-1 AFHC\cite{ts1}. Again shifting the momentum as $k=\pi + q$, the effective Hamiltonian is given by
    
\begin{eqnarray}
\label{eq:mfham2}
\H &=&  \sum_{q}\Delta\{\frac{\xi^2q^2}{2}-(hs-1)\}\alpha^{\dagger}_{q,s} \alpha_{q,s} \nonumber \\
&+&\frac{1}{N}\sum_{q_1,q_2,q_3,q_4,s,s'}c_{s,s'}\alpha^{\dagger}_{q_1,s}  \alpha^{\dagger}_{q_2,s'} \nonumber \\
&\times&\alpha_{q_3,s'}\alpha_{q_4,s}\delta_{q_1+q_2,q_3+q_4},
\end{eqnarray}
where $h= H/H_{\mbox{c}}$. The suffices $s$ and $s'$ denote the $z$-component of the spin of the excitations and take the values 1, 0 and $-1$. The values of $\Delta$ and $\xi$ are estimated by extrapolating the exact diagonalization data\cite{kh1,khd} to $N \rightarrow \infty$ by the Shanks' transformation\cite{ds1} as shown in Fig. \ref{fig1}.  Our boson Hamiltonian again yields the corresponding boson Hamiltonian for the $S=1$ AFHC \cite{ts1} in the limit $\beta \rightarrow -\infty$.

The magnetization per unit cell is given by 
\begin{equation}
M= \frac{1}{N}\sum_q \{<\alpha_{q,1}^{\dagger}\alpha_{q,1}> - <\alpha_{q,-1}^{\dagger}\alpha_{q,-1}>\}.
\end{equation}

We further neglect the interaction $c_{s,s'}$ for $s \neq s'$. This is justified as follows. In the limit $h << 1$, and $T << \Delta$, the occupation number of each bosonic state is low and all interactions may be neglected.  On the other hand, in the vicinity of the lower critical field ($h \sim 1$), the occupation numbers of the low energy states of $s=1$ excitation are not necessarily small. Therefore the interaction $c_{1,1}$ must not be neglected. However, other interactions can be neglected because the occupation numbers of other excitations are small at low temperatures. Nevertheless we keep the interaction $c_{-1,-1}$ for the sake of symmetry between the excitations with $s=1$ and $s=-1$ . Following Takahashi and Sakai \cite{ts1}, the value of the interaction constant $c_{s,s}$ is identified as $c_{s,s}=1/2$  by comparing the ground state magnetization for $H_{\mbox{s}} >> H >> H_{\mbox{c}}$ obtained from the bosonic theory with the classically calculated magnetization $M = H$. Thus both $s=1$ and $s=-1$ excitations are described as $\delta$-function bosons and the magnetization curve at finite temperature is obtained using the numerical solution of the integral equations (\ref{eq:eps}) and (\ref{eq:rho}) with
\begin{equation}
A=\frac{2(hs-1)}{\xi^2}, \ \ \ c=\frac{1}{\Delta\xi^2},\ \ \ t=\frac{2T}{\Delta\xi^2},
\end{equation}
for $s = \pm 1$. The magnetization per unit cell is given by
\begin{equation}
M = \int_{-\infty}^{\infty}\rho_{1}(k)dk - \int_{-\infty}^{\infty}\rho_{-1}(k)dk,
\end{equation}
where $\rho_s$ is the solution of Eq. (\ref{eq:rho}) for each value of $s$.

The equations (\ref{eq:eps}) and (\ref{eq:rho}) are solved numerically. Figures \ref{fig2}(a-f) shows the magnetization curves for $\beta = -0.5, -1 $ and $-3$. The open circles and open squares represent the results of high field and low field approximations, respectively.

\section{Quantum Monte Carlo Simulation}

We have also performed the quantum Monte Carlo simulation for the Hamiltonian (\ref{eq:ham1}) to check the region of applicability of our boson approximation. The algorithm is the standard world line method using the checker board decomposition with local and global flip in the spatial and Trotter direction. The extrapolation with respect to the Trotter number is performed in the following way. For $T=0.1$ and $\beta=-3$, for which the worst convergence is expected, we have taken the Trotter numbers $L=10, 12$ and 18 and checked that the correction term is almost proportional to $1/L^2$ as expected. The actual extrapolation is made by extrapolating the data for $L=12$ and 18 linearly to $1/L^2$. The system size $N$ is fixed to $N=16$ (32 spins). The measurement is done over typically $4 \times 10^6$ Monte Carlo steps. The results are shown in Figs. \ref{fig2} (a-f) by filled circles. Typical error bars are smaller than the size of the symbols.
\section{Results and Discussion}

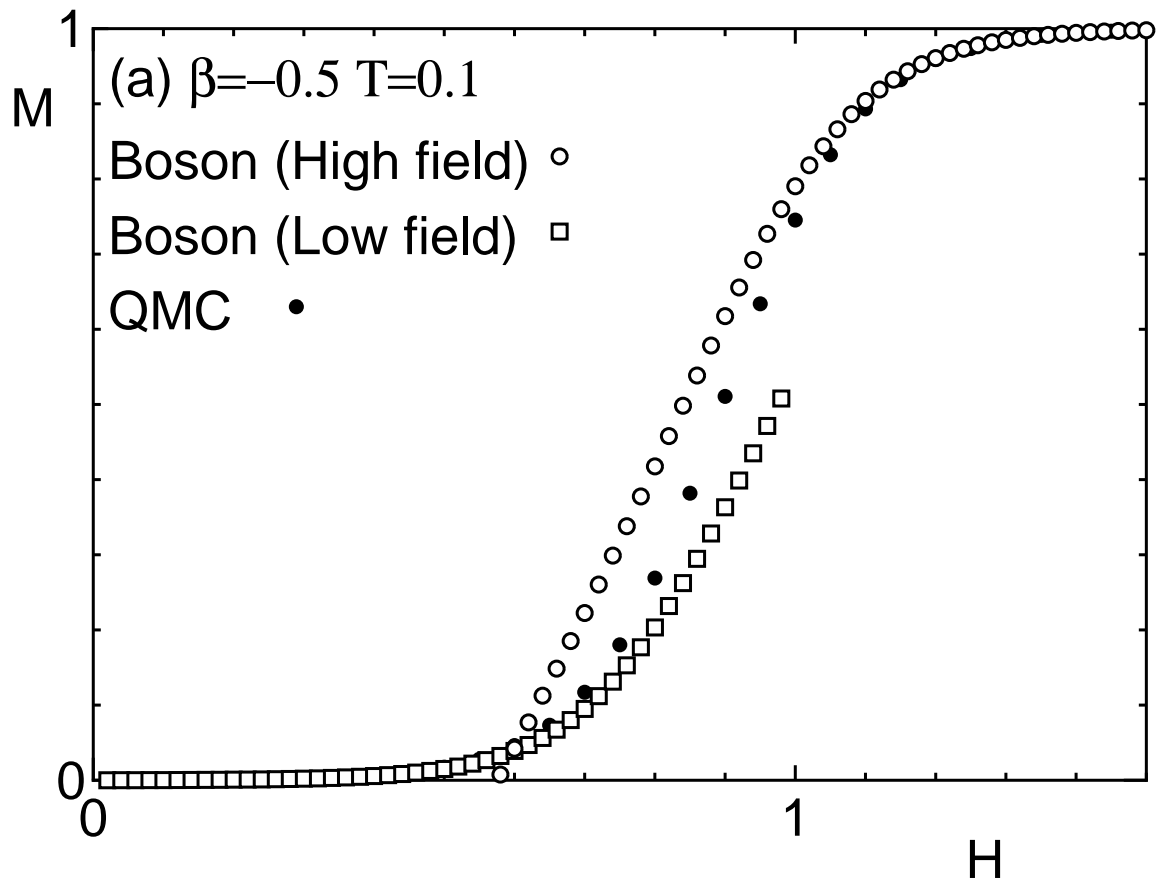
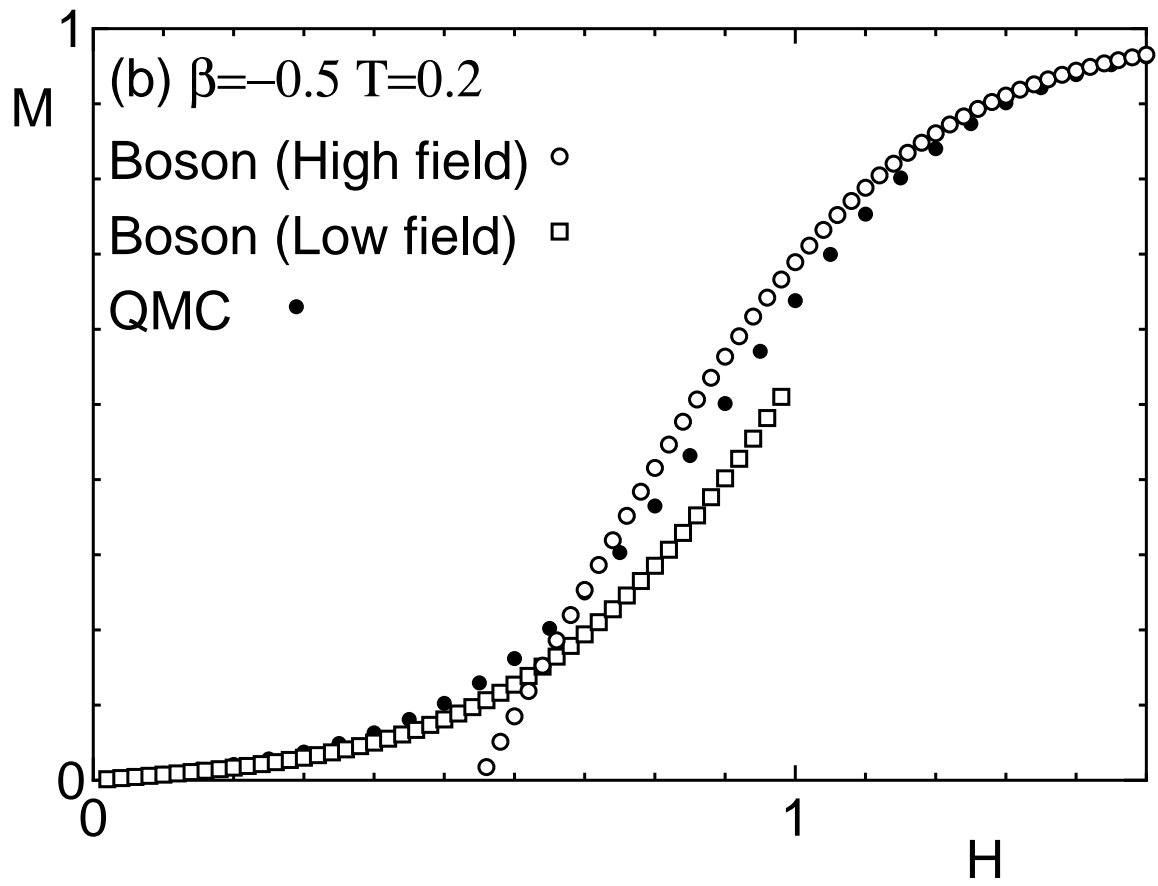
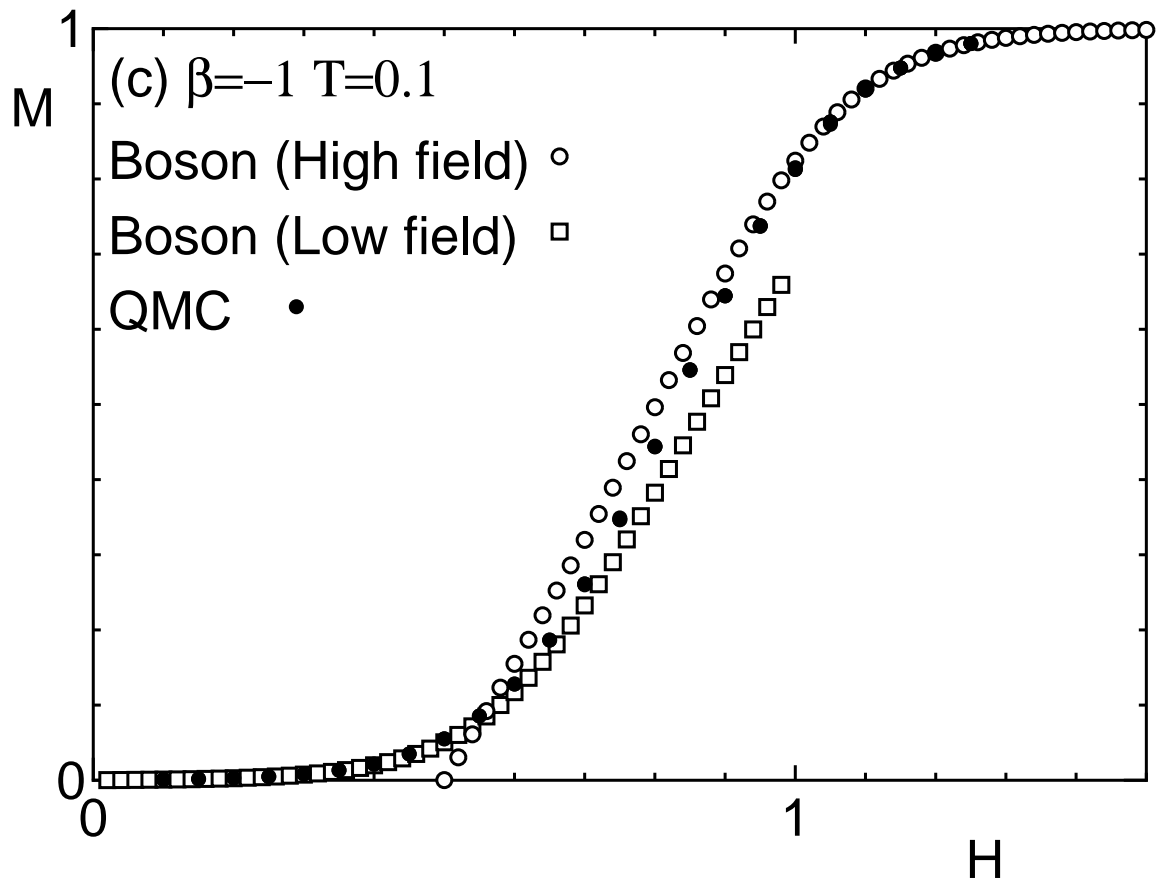
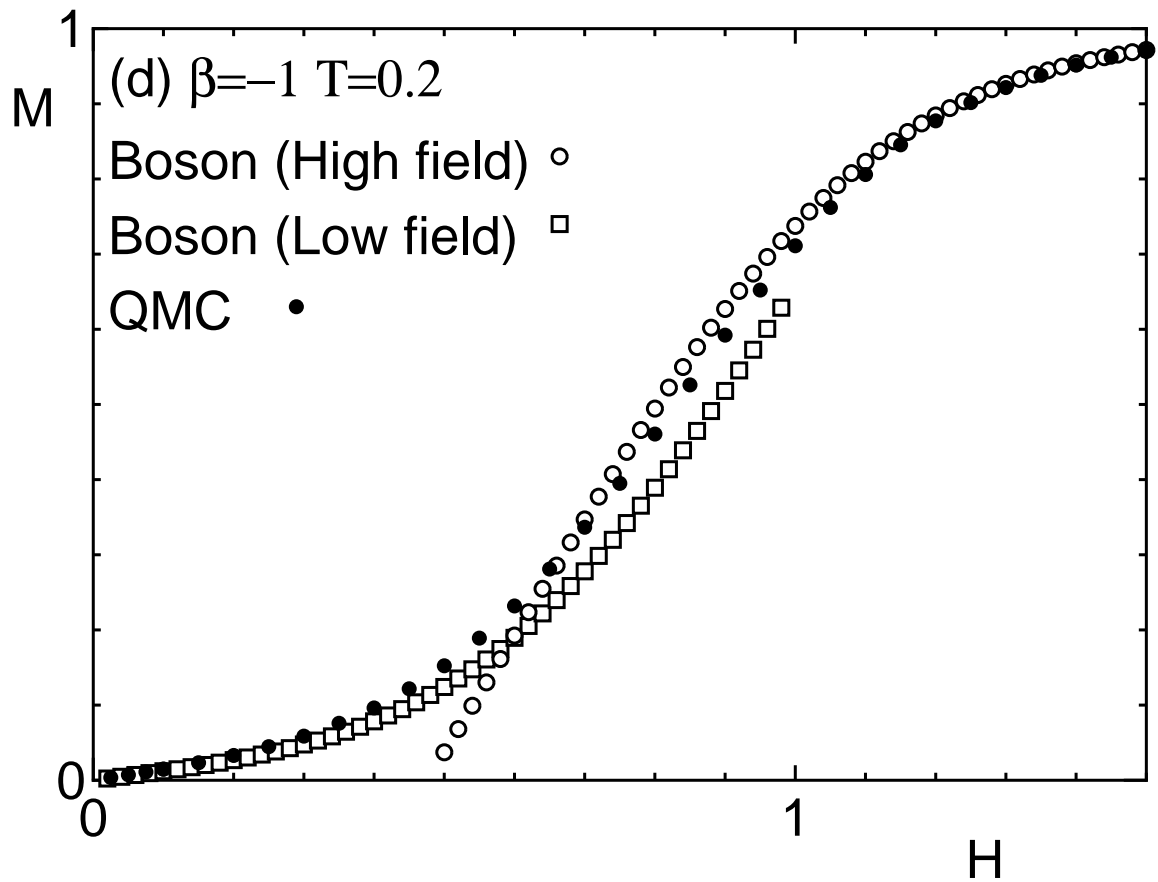
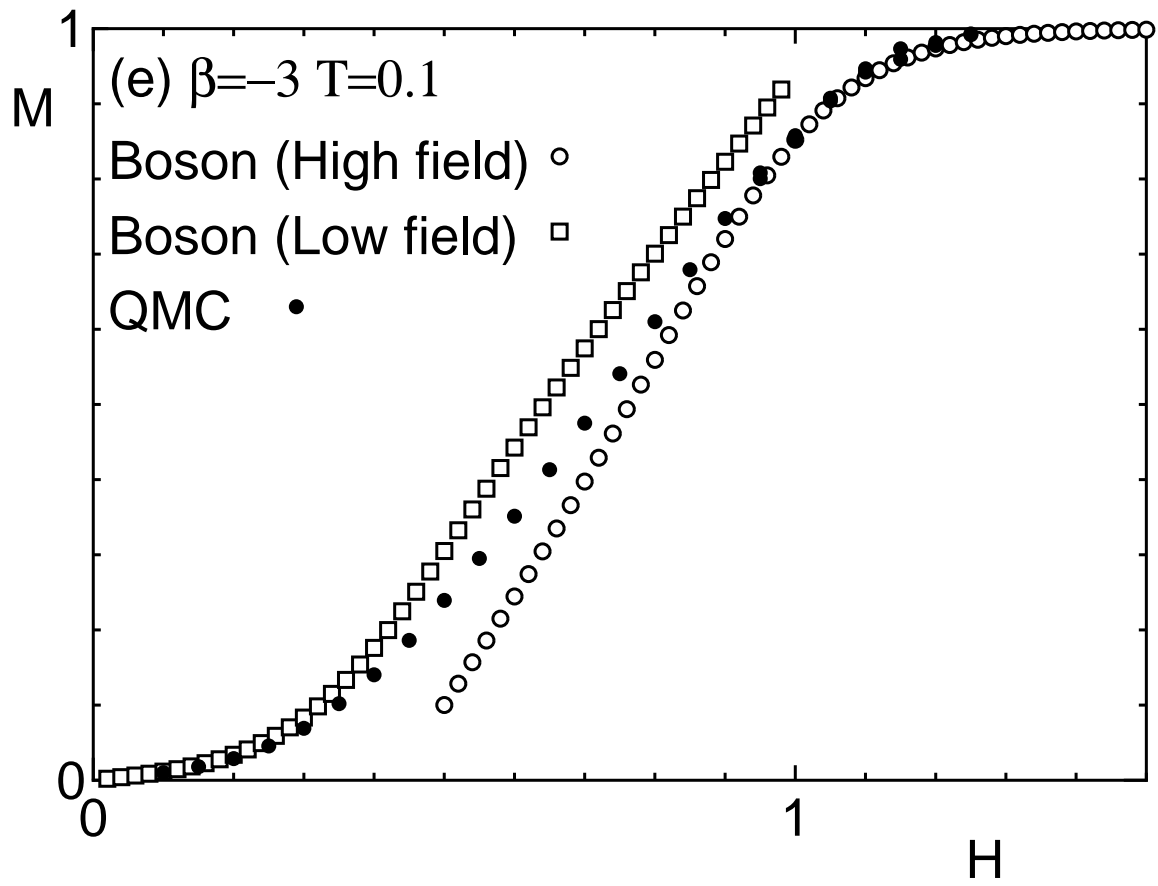
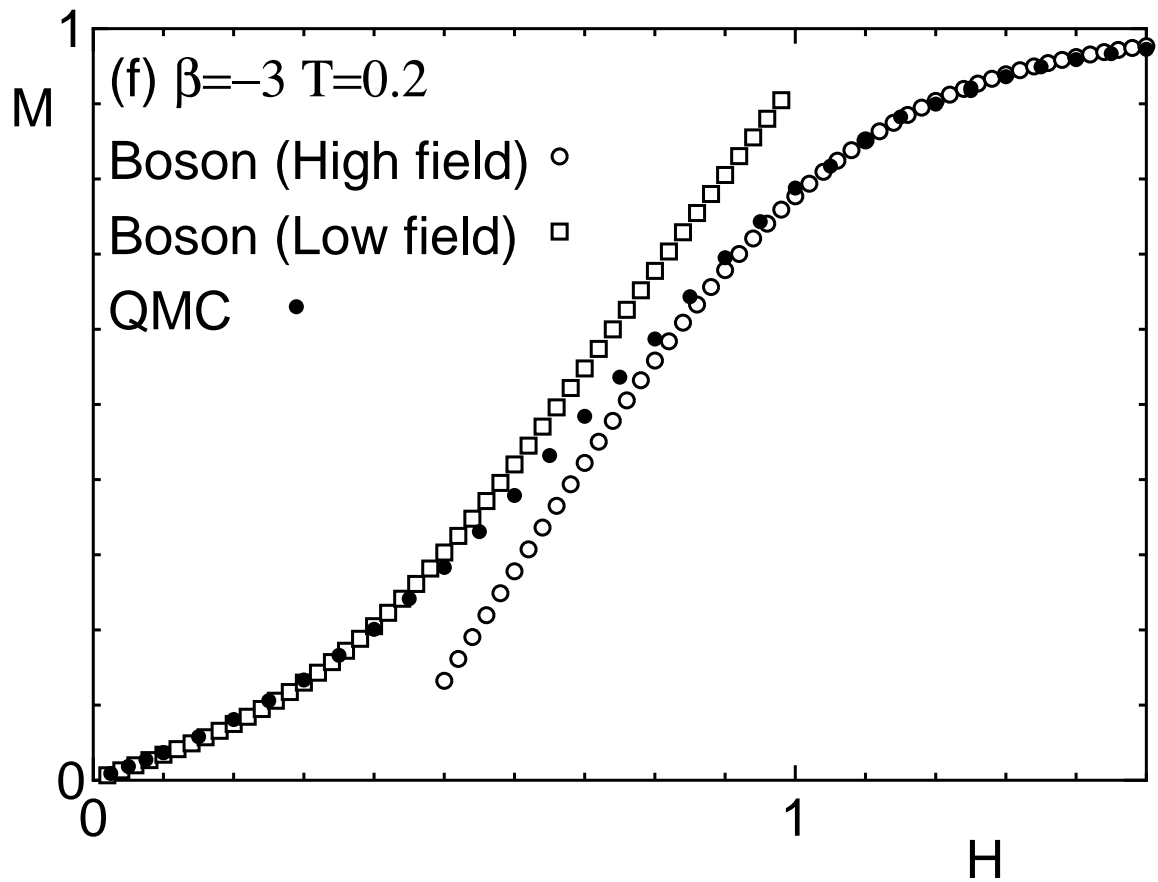
\begin{figure}
\caption{The magnetization curves of the present model. The results of the Monte Carlo simulation, high field approximation and low field approximation are shown by the filled circles, open circles and open squares, respectively. The values of temperature $T$ and $\beta$ are shown in the figure.}\label{fig2}
\end{figure}

The comparison of the data shows that the boson approximation is fairly good for low field region $0 < M \simleq  0.1$ and near the saturation field $0.9 \simleq M < 1$. In the intermediate regime, however, there exist considerable discrepancy between our approximation and numerical results. One of the origin of the discrepancy is attributed to the finite band width of magnons. If we approximate them by the unbounded parabolic bands as Eqs. (\ref{eq:mfham1}) and (\ref{eq:mfham2}), we overestimate the number of magnons with large $\mid q \mid $ in the limit of high (low) magnetic field within the low (high) field approximation. In the low field approximation, however, for the intermediate magnetic field, the number of magnons with $q \sim \pi$ are {\it underestimated} because the true dispersion is convex around this point\cite{khd} and these excitations are more easily excited than the parabolic case. For large negative $\beta$, the convex part of the dispersion is absorbed in the two magnon continuum\cite{khd}  and the latter effect becomes less pronounced. It should be also noted that the multimagnon effect which is not correctly treated within the present approximation would also cause the discrepancy. For the high field approximation, the $-$ branch comes down as $\mid \beta \mid$ decreases and it touches the $+$ branch for $\beta = -1$. Therefore the contribution from the $-$ branch is no more negligible for  $\beta \sim -1$. This causes the underestimation of the magnon number for small $\beta$ in the present single mode approximation. These arguments apply not only for the magnons in the ground state but also to the thermally excited magnons. Actually, the qualitative tendency of deviation does not depend on temperature as shown in Fig. \ref{fig2}.

As an alternative method to map the spin chain onto the $\delta$-function boson gas in the high field regime, Akutsu, Okunishi and Hieida\cite{aoh1} have proposed to use the 2-body $S$-matrix. They applied this method to the $S=1$ bilinear-biquadratic model and determined the interaction parameter $c$ of the $\delta$-function bose gas so that its 2-body $S$-matrix coincides with that of the  spin chain in the long wave length limit. Surprisingly, their expression for $c$ turned out to be different from the one obtained by means of the  Dyson-Maleev transformation\cite{ts1} and diverges as the biquadratic term vanishes. Although the origin of the discrepancy is unclear, it may be related with the order in which the long wavelength limit is taken and mapping onto the boson gas is made. Actually, in the present model, the 2-body $S$-matrix cannot be represented by a local phase shift and it is difficult to construct the 2-body $S$-matrix before taking the long wave length limit. This problem is left for the future study.

From the experimental side, several materials which realizes the spin-1/2 ferromagnetic-antiferromagnetic alternating chain are synthesized. The magnetization curves are measured for $[\mbox{Cu(TIM)]CuCl}_4$\cite{hag1}, $\mbox{(4-BzpipdH)CuCl}_3$\cite{hag1} and methyl nitronyl nitroxide\cite{tak1}. The direct comparison of the present calculation and the experimental data is difficult, however, because of the presence of the interchain interaction which not only leads to the  antiferromagnetic ordering at fairly high temperatures but also has influence on the magnetization curves in the paramagnetic phase. This kind of interaction cannot be treated within the present scheme which makes use of the Bethe ansatz solution. In this context, the synthesis of similar compounds with reduced interchain interaction is hoped for the direct comparison of the present calculation and experiment. On the other hand, the inclusion of other kinds of intrachain interaction is straightforward in the present scheme. Therefore our method gives a convenient way to calculate the low temperature magnetization curves of one-dimensional antiferromagnets in the low field and high field regime semi-quantitatively without resorting to elaborate numerical simulations.

The author would like to thank S. Miyashita for providing the source code of the quantum Monte Carlo simulation program. He also thanks M. Hagiwara for showing him his experimental data prior to publication and for stimulating discussion. Thanks are also due to Y. Natsume and T. Suzuki for sending their Monte Carlo data prior to publication and for invaluable comments. The numerical simulation is performed using the FACOM VPP500 at the Supercomputer Center, Institute for Solid State Physics, University of Tokyo, the HITAC S820/80 and SR2201 at the Information Processing Center of Saitama University.  This work is supported by the Grant-in-Aid for Scientific Research from the Ministry of Education, Science, Sports and Culture.

%\vspace{1080pt}

\end{document}